\documentclass[a4paper]{article}

\usepackage{amsfonts,amssymb,amsmath,graphicx}
\usepackage[a4paper,left=25mm,right=25mm,top=25mm,bottom=30mm]{geometry}
\usepackage[unicode,colorlinks=true,urlcolor=black,linkcolor=black,citecolor=black,pdfborder={0 0 0}]{hyperref}
\usepackage[auth-lg,affil-it]{authblk}
\usepackage{cite}

\begin{document}

\title{Simulating Thin Sheets:\\Buckling, Wrinkling, Folding and Growth}

\author[1]{Roman Vetter}
\author[1,2]{Norbert Stoop}
\author[1]{Falk K.~Wittel}
\author[1]{Hans J.~Herrmann}

\affil[1]{Computational Physics for Engineering Materials, IfB, ETH Zurich, Schafmattstrasse~6, CH-8093~Zurich, Switzerland}
\affil[2]{IAS Institute of Applied Simulations, ZHAW Zurich University of Applied Sciences, CH-8820~W\"adenswil, Switzerland}

\date{11 March 2014}

\maketitle

\begin{abstract}
Numerical simulations of thin sheets undergoing large deformations are computationally challenging. Depending on the scenario, they may spontaneously buckle, wrinkle, fold, or crumple. Nature's thin tissues often experience significant anisotropic growth, which can act as the driving force for such instabilities. We use a recently developed finite element model to simulate the rich variety of nonlinear responses of Kirchhoff--Love sheets. The model uses subdivision surface shape functions in order to guarantee convergence of the method, and to allow a finite element description of anisotropically growing sheets in the classical Rayleigh--Ritz formalism. We illustrate the great potential in this approach by simulating the inflation of airbags, the buckling of a stretched cylinder, as well as the formation and scaling of wrinkles at free boundaries of growing sheets. Finally, we compare the folding of spatially confined sheets subject to growth and shrinking confinement to find that the two processes are equivalent.
\end{abstract}

\section{Introduction}

Thin sheets are omnipresent in nature, technology and everyday life, appearing at virtually all length scales. Being much thinner in one than in the other two dimensions, they can develop an unparalleled, rich variety of deformation modes when subjected to external forces, spatial constraints, or intrinsic growth. They buckle, wrinkle, fold, and crumple. Numerical simulations are an often-indispensable approach for studying the complex interplay of these modes. The folding and crumpling of a piece of paper \cite{LGLMW95,BAP97,BK05,VG06,TAT08,TAT09,VG11} and metal sheets wrinkling and crumpling in vehicle collisions \cite{O98,WWK01,MMIKH01,MO05} are two out of many examples. For many such problems, the finite element method (FEM) has shown to be amongst the most efficient and flexible tools, especially in cases with strong material nonlinearity, complex geometry, or anisotropy. Even though the Kirchhoff--Love theory \cite{L88} provides a simple kinematic description, numerically sound finite element implementations of thin sheets have turned out to be difficult and cumbersome in the past. These problems can be successfully overcome since the subdivision surface paradigm was introduced to the FEM \cite{COS00,CO01}.

Large deformations of soft thin tissue such as insect wings, plant leaves, cell membranes, or flowers are often induced by growth (or shrinkage) \cite{T95,DBA08}, inevitably leading to the development of residual stress \cite{H86,SZJNH96}. In this paper, we present an extension of the Kirchhoff--Love theory to allow for anisotropic in-plane growth, which we implement with Loop subdivision shape functions. The combination of these two concepts grants access to a very straightforward and highly efficient, yet powerful and flexible numerical tool for the simulation of nonlinear thin sheet mechanics. Our approach accounts for the change of reference curvature when the surface grows, generalizing recently developed tethered mass-spring models \cite{MSSR03,MDS07}. The next section summarizes the mentioned extension. It is followed in the subsequent sections by a series of thin sheet problems that we solve using the developed FEM implementation. Special attention is paid to the formation of self-similar wrinkles along a plastically stretched free edge as well as on the scaling of single-wavelength wrinkles of growing cylinders similar to flowers.

\section{The Kirchhoff--Love Sheet with Anisotropic Growth}

\begin{figure}[!hbp]
	\centering
	\includegraphics{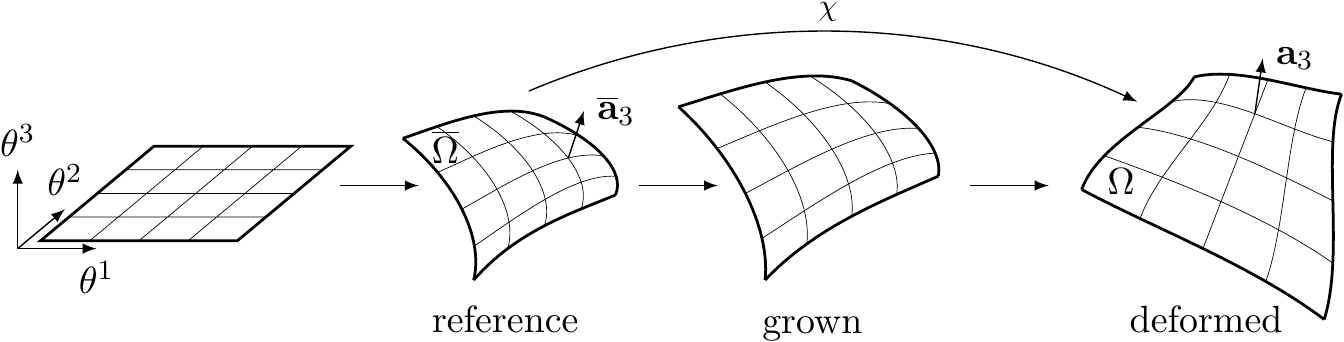}
	\caption{Reference, grown and deformed configurations of the sheet's middle surface.}
	\label{fig:coord_sys}
\end{figure}

Let $\overline{\Omega} \subset \mathbb{E}^3$ be the stress-free undeformed (``reference'') middle surface of a sheet with small thickness $h$. Under the action of external forces or growth, the sheet deforms into a new configuration with middle surface $\Omega \subset \mathbb{E}^3$. In the following, let Greek indices $\alpha,\beta,\gamma,\delta\in\{1,2\}$, and Latin indices $i,j\in\{1,2,3\}$. Lower (upper) indices will denote covariant (contravariant) components. Moreover, let $\{\theta^1,\theta^2, \theta^3\}$ be a curvilinear coordinate system, and let $\overline{{\bf x}}(\theta^1,\theta^2)$ and ${\bf x}(\theta^1,\theta^2)$ be parametrizations of $\overline{\Omega}$ and $\Omega$, respectively (see Fig.~\ref{fig:coord_sys}). The material points $\overline{{\bf p}}$ and ${\bf p}=\chi(\overline{{\bf p}})$ in the reference and deformed sheet are parametrized as
\begin{equation}
\overline{{\bf p}}(\theta^1,\theta^2,\theta^3) = \overline{{\bf x}}(\theta^1,\theta^2) + \theta^3 \overline{{\bf a}}_3(\theta^1,\theta^2) \quad \mathrm{and} \quad {\bf p}(\theta^1,\theta^2,\theta^3) = {\bf x}(\theta^1,\theta^2) + \theta^3 {\bf a}_3(\theta^1,\theta^2),
\label{eq:r_def}
\end{equation}
where $\theta^3 \in [-h/2,h/2]$. $\chi$ is a diffeomorphism that maps from the reference to the deformed material positions. The tangent spaces of $\overline{\Omega}$ and $\Omega$ are spanned by the respective vector fields
\begin{equation}
\overline{{\bf a}}_{\alpha}(\theta^1,\theta^2) = \overline{{\bf x}}_{,\alpha} = \frac{\partial\overline{\bf x}}{\partial\theta^{\alpha}} \quad \mathrm{and} \quad {\bf a}_{\alpha}(\theta^1,\theta^2) = {\bf x}_{,\alpha} = \frac{\partial\bf x}{\partial\theta^{\alpha}}.
\end{equation}
By virtue of the Kirchhoff assumption, straight material lines normal to the middle surface retain these properties as well as their length. They are determined by the unit normal vectors
\begin{equation}
\label{eq:conda3}
{\overline{\bf a}}_3 = \frac{{\overline{\bf a}}_1\times {\overline{\bf a}}_2}{|{\overline{\bf a}}_1\times {\overline{\bf a}}_2|} \quad \mathrm{and} \quad {\bf a}_3 = \frac{{\bf a}_1\times {\bf a}_2}{|{\bf a}_1\times {\bf a}_2|}.
\end{equation}

\noindent The covariant components of the first fundamental forms follow as
\begin{equation}
\overline{a}_{\alpha\beta} = \overline{{\bf a}}_{\alpha} \cdot \overline{{\bf a}}_{\beta} \quad \mathrm{and} \quad a_{\alpha\beta} = {\bf a}_{\alpha} \cdot {\bf a}_{\beta},
\end{equation}
while those of the second fundamental forms are given by
\begin{equation}
\overline{b}_{\alpha\beta} = \overline{{\bf a}}_{3} \cdot \overline{{\bf a}}_{\alpha,\beta} \quad \mathrm{and} \quad b_{\alpha\beta} = {\bf a}_{3} \cdot {\bf a}_{\alpha,\beta}.
\end{equation}
Assuming that the thin sheet obeys the St.~Venant--Kirchhoff law of linear elasticity, the connection between its kinematics and energetics is provided by the Koiter energy density functional \cite{K66,C00}. Let the sheet be characterized by Young's modulus $E$ and Poisson's ratio $\nu$. The elastic energy $U_{\mathrm{e}}$ of the Koiter sheet is obtained by integrating the energy per unit area over the middle surface:
\begin{equation}
\label{eq:elasticenergy}
U_{\mathrm{e}}[\overline{\bf x},{\bf x}] = \frac{1}{2} \int_{\overline{\Omega}} \frac{Eh}{1-\nu^2} \left(H^{\alpha\beta\gamma\delta}\alpha_{\alpha\beta} \alpha_{\gamma\delta} + \frac{h^2}{12} H^{\alpha\beta\gamma\delta}\beta_{\alpha\beta} \beta_{\gamma\delta}\right)\,\mathrm{d}\overline{\Omega},
\end{equation}
where $\mathrm{d}\overline{\Omega} = |\overline{{\bf a}}_1 \times \overline{{\bf a}}_2|\;\mathrm{d}\theta^1 \mathrm{d}\theta^2$. The Einstein summation applies to repeated indices. $H$ is often referred to as the ``elastic tensor'', and is given component-wise by
\begin{align}
H^{\alpha\beta\gamma\delta} = \nu\overline{a}^{\alpha\beta}\overline{a}^{\gamma\delta}+\frac{1-\nu}{2}(\overline{a}^{\alpha\gamma}\overline{a}^{\beta\delta}+\overline{a}^{\alpha\delta}\overline{a}^{\beta\gamma}).
\end{align}
$\alpha=(a-\overline{a})/2$ and $\beta=\overline{b}-b$ are the in-plane ($2\!\times\!2$) membrane and bending strain tensors, respectively. The Koiter shell (\ref{eq:elasticenergy}) can be extended to incorporate anisotropic growth through the multiplicative decomposition of the geometric deformation gradient $\nabla\chi = {\bf F}_{\mathrm e}\,{\bf F}_{\mathrm g}$ \cite{L69,RHM94} into a growth tensor ${\bf F}_{\mathrm g}$ and a purely elastic response ${\bf F}_{\mathrm e}$, that ensures continuity and compatibility of the body. Owing to the Kirchhoff constraints, we may write
\begin{equation}
{\bf F}_{\mathrm g} = \begin{bmatrix}G & 0\\0^{\mathrm{T}} & 1\end{bmatrix}\qquad(G\in\mathbb{R}^{2\times2}\;\mathrm{symmetric}),
\end{equation}
and the growth-modified elastic strains then simply read \cite{VSJWH13}
\begin{align}
\alpha &= \frac{1}{2}\left(G^{-\mathrm{T}}aG^{-1} - \overline{a}\right),\\
\beta &= \overline{b} - G^{-\mathrm{T}}bG^{-1}.
\end{align}
We further augment the elastic energy (\ref{eq:elasticenergy}) with an inertial term to capture the dynamics of the thin sheet. The kinetic energy reads
\begin{equation}
\label{eq:kineticenergy}
U_{\mathrm{k}}[\overline{\bf x},{\bf x}] = \frac{1}{2} \int_{\overline{\Omega}} h\rho\, \dot{{\bf x}} \cdot \dot{\bf x}\,\mathrm{d}\overline{\Omega},
\end{equation}
where $\rho$ is the mass density of the sheet, and $\dot{\bf x}=\partial{\bf x}/\partial t$ is the velocity. Our aim is to find the minimizer {\bf x} of the total energy $U=U_{\mathrm{e}}+U_{\mathrm{k}}$ for given growth tensors ${\bf F}_{\mathrm g}$ or external driving forces.

\section{Finite Element Implementation with Subdivision Surfaces}

To account for out-of-plane bending rigidity, the bending term in Eq.~(\ref{eq:elasticenergy}) integrates the Gaussian and mean curvatures, which comprise second derivatives of the displacement field ${\bf u}={\bf x}-\overline{\bf x}$, over the middle surface. For boundedness of the integral in the weak formulation, continuously differentiable finite element shape functions ($C^1$-continuity) are needed. This requirement has proven very challenging in the history of shell finite elements, until Cirak et al.~\cite{COS00,CO01} have introduced Loop subdivision surfaces to the FEM. A fundamental difference to traditional finite elements is that subdivision surfaces gain $C^1$-continuity at the expense of a larger support of the individual shape functions. Details on their implementation are given in Refs.~\cite{COS00,VSJWH13}.

Aside from guaranteeing convergence, subdivision surfaces allow a classical Rayleigh--Ritz formulation of the sought finite element deformation: no rotational variables are needed and the only unknowns are the three nodal displacements. Moreover, a single quadrature point per element is sufficient \cite{COS00,CO01,VSJWH13}, rendering this finite element approach computationally highly efficient and flexible. Of course, increasing the number of quadrature points may assist in resolving strongly anisotropic growth fields or constitutive relations. We employ Loop subdivision surface shape functions here to minimize the total energy $U$ numerically by solving Newton's equation of motion with a standard predictor-corrector scheme. Subcritical viscous damping is added for numerical stability and equilibration.

\section{Inflated Pillows and Airbags}

As a first instance of folding and wrinkling, we reproduce the inflation of pillows and airbags from Ref.~\cite{CO01}. A square sheet with diagonal length $d=120$ cm and thickness $h=1$ mm and a circular sheet with radius $R=35$ cm and thickness $h=0.4$ mm are instantaneously pressurized with 5 kPa to buckle out of their flat initial configuration. The elastic moduli are given by $E=588$ MPa, $\nu=0.4$ and $E=60$ MPa, $\nu=0.3$, respectively. We exploit the reflection symmetry by simulating only the upper half of the geometry. The quasi-static equilibrium configurations are shown in Fig.~\ref{fig:airbags}, and a movie of the dynamic deformation is provided in the supplementary material. The square pillow features a distinct folding pattern in the middle of the four edges, resulting from the non-uniform distribution of Gaussian curvature: The edges are closer to the point of maximal uplift than the corners, thus getting curved more and pulled towards the center. They are laterally stretched and longitudinally compressed, yielding the observed folds. The circular airbag, on the other hand, is initially axisymmetric and therefore behaves differently. Wrinkles develop similarly to elastic plates stamped into curved cavities \cite{HBB12} or ultrathin films placed on fluid drops \cite{KSDM12}. On top of these low-amplitude wrinkles, axisymmetry of the stress field is broken and large crumples occur that localize the geometrically imposed Gaussian curvature.

\begin{figure}[!htbp]
	\centering
	\includegraphics{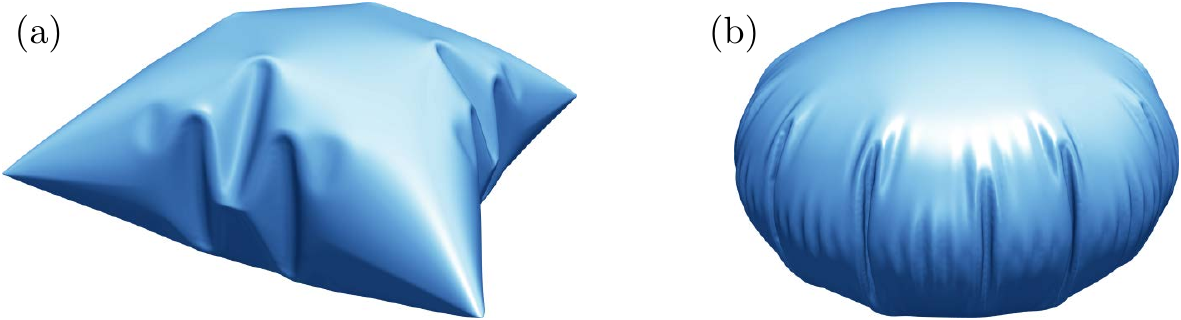}
	\caption{Inflated square pillow (a) and circular airbag (b).}
	\label{fig:airbags}
\end{figure}

\section{Buckling of a Stretched Cylinder}

\begin{figure}[!htbp]
	\centering
	\includegraphics{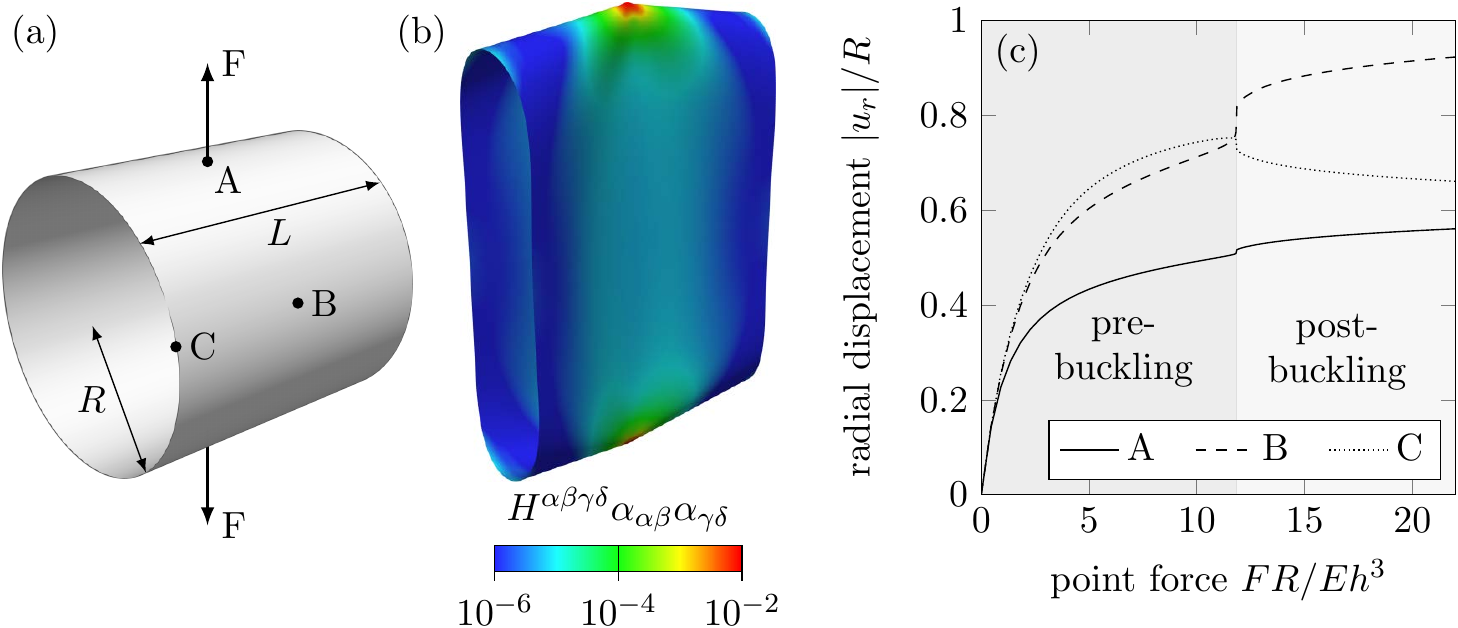}
	\caption{(a) Reference state of the cylindrical sheet. (b) Equilibrium solution at $F=1.92F_\mathrm{c}$. The stretching energy density is shown on a logarithmic color scale. (c) Normalized radial displacements vs.~the rescaled point force.}
	\label{fig:stretched_cylinder}
\end{figure}

A frequently studied buckling problem is the laterally stretched open cylinder \cite{NGMA09,W10}. Two opposite point forces of equal increasing magnitude $F$ cause the cylindrical sheet with radius $R=4.953$ cm, length $L=10.35$ cm, thickness $h=0.94$ mm, and free edges to first bend before snapping through at $F_\mathrm{c}=11.836\,Eh^3/R$ to the post-buckling regime, where further deformations are dominated by stretching. The elastic moduli are fixed to $E=10.5$ MPa, $\nu=5/16$. In Fig.~\ref{fig:stretched_cylinder}, we plot the radial displacements of the points A, B and C, which are degenerate at the buckling threshold $F_\mathrm{c}$. The corresponding movie can be found in the supplementary material.

\section{Boundary Instabilities and Wrinkling}

\begin{figure}[!htbp]
	\centering
	\includegraphics[width=\textwidth]{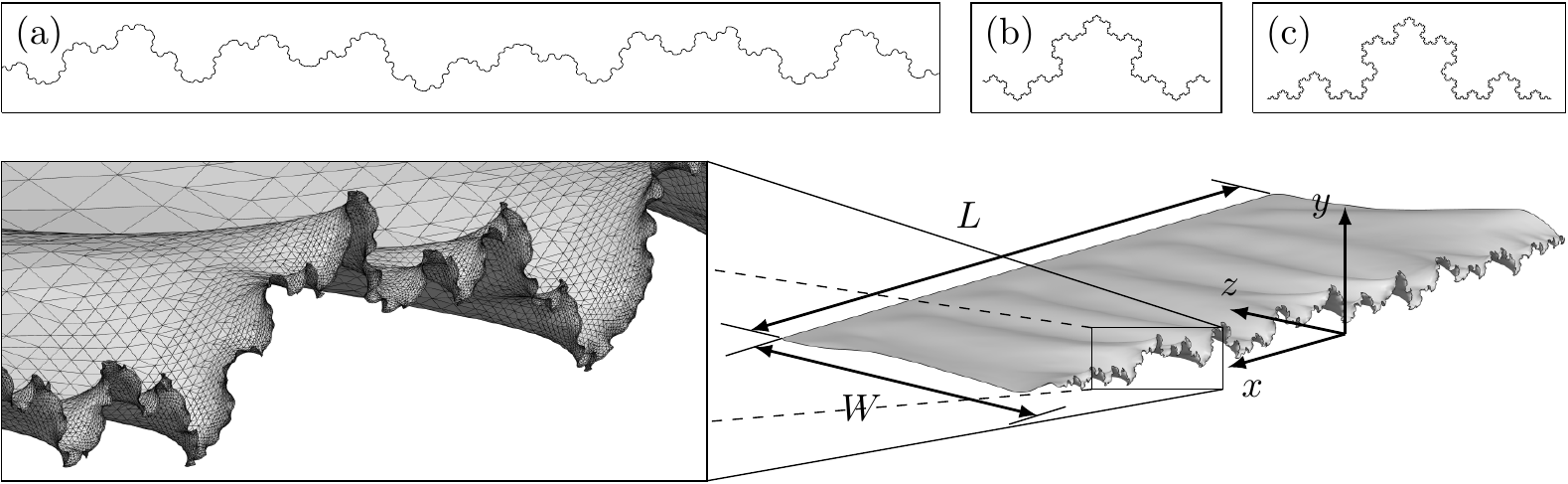}
	\caption{Self-similar sheet boundary after growing according to Eq.~(\ref{eq:growth_tensor}). (a) Projection of the grown edge onto the $xy$ plane. (b) $60^{\circ}$ Fibonacci word fractal. (c) Koch snowflake.}
	\label{fig:selfsim_growth}
\end{figure}

An interesting feature observed in plant growth is the occurrence of self-similar wrinkles along free tissue boundaries, such as the edges of flowers and leaves \cite{SRMSS02,M03,MSSR03,A04,MDS07,SRS07}. The morphology is very similar to the shape of torn plastic sheets, and apparently, both phenomena are characterized by a plastic longitudinal metric profile $g_l(z) = 1/(1+z/l)$, where $l>0$ is a characteristic length scale and $z\geq 0$ is the coordinate perpendicular to the growing edge. Audoly and Boudaud \cite{AB03} were able to show that the solution of the F\"oppl--von K\'arm\'an equations on the edge of a free rectangular sheet with such growth profiles consists of self-similar wrinkles governed by odd integral scaling factors. The F\"oppl--von K\'arm\'an equations are, however, geometrically limited as they don't allow reentrancy. The present growing Koiter shell model allows us to numerically solve the problem with its full geometric nonlinearity taken into account. Consider a flat rectangular sheet of thickness $h=10^{-4}$, length $L=4$, and width $W=1$, initially lying in the $xz$ plane. We clamp the long edge at $z=W$ and constrain the short edges to stay at $x=-L/2,L/2$, leaving them free to move in other directions. Plastic growth is imposed by setting the growth tensor to
\begin{equation}
{\bf F}_{\mathrm g} = \mathrm{diag}\big(1+g_l(z),1,1\big)
\label{eq:growth_tensor}
\end{equation}
in Cartesian coordinates $(x,y,z)$, and we choose a characteristic length $l=40h$ for the growth field in this example. Fig.~\ref{fig:selfsim_growth} shows the resulting equilibrated configuration after growth (or tearing), and a movie showing the equilibration is provided in the supplementary material. The self-similarity of the free edge at $z=0$ is apparent, clearly resembling the wrinkling cascades observed in experiments \cite{SRMSS02,MDS07}.

\begin{figure}[!htbp]
	\centering
	\includegraphics{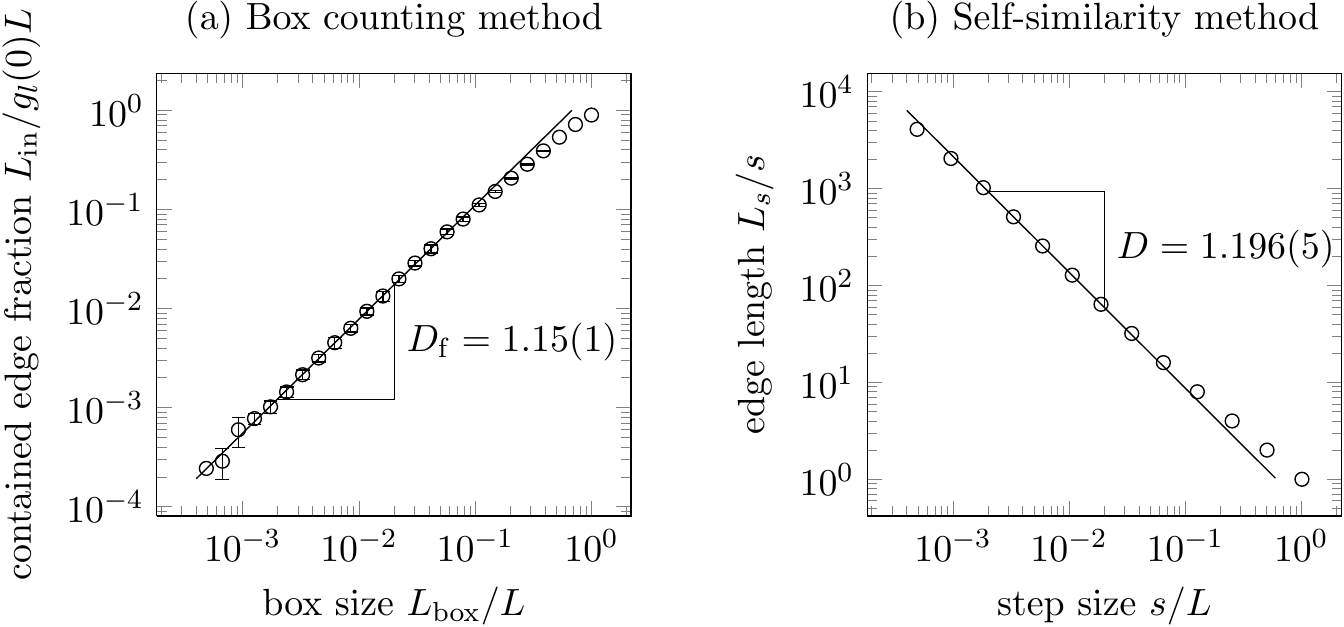}
	\caption{Fractal dimension of the edge at $z=0$ of a thin sheet grown according to Eq.~(\ref{eq:growth_tensor}), measured with two standard methods.}
	\label{fig:fracdim}
\end{figure}

We have measured the fractal dimension of the grown edge depicted in Fig.~\ref{fig:selfsim_growth}(a) using the box counting method \cite{FW79} and the self-similarity method \cite{M67}. In the former, the curve length $L_{\mathrm{in}}$ contained in a cubic box is determined as a function of the edge length $L_{\mathrm{box}}$ of the box. A fractal curve is expected to scale as $L_{\mathrm{in}}\sim L_{\mathrm{box}}^{D_{\mathrm f}}$. Such scaling is indeed observed with fractal dimension ${D_{\mathrm f}}=1.15(1)$ (Fig.~\ref{fig:fracdim}(a)). The scaling breaks down due to the influence of the clamped opposite edge, which introduces a global straightening effect when the box size is large ($L_{\mathrm{box}}\approx W$). The second method is more robust to global orientation and is thus better suited here. The length $L_s$ of a piecewise linear path along the curve with segment size $s$ is measured and expected to scale as $L_s\sim s^{1-D}$. We find a self-similarity dimension $D=1.196(5)$ (Fig.~\ref{fig:fracdim}(b)), which is very close to the Hausdorff dimension of a 60-degree Fibonacci word fractal (Fig.~\ref{fig:selfsim_growth}(b)), $D_{\mathrm H}=1.2083$ \cite{MD09}, and a bit lower than that of a triadic Koch curve (Fig.~\ref{fig:selfsim_growth}(c)), $D_{\mathrm H}=1.2619$ \cite{M83}.

\begin{figure}[!htbp]
	\centering
	\includegraphics{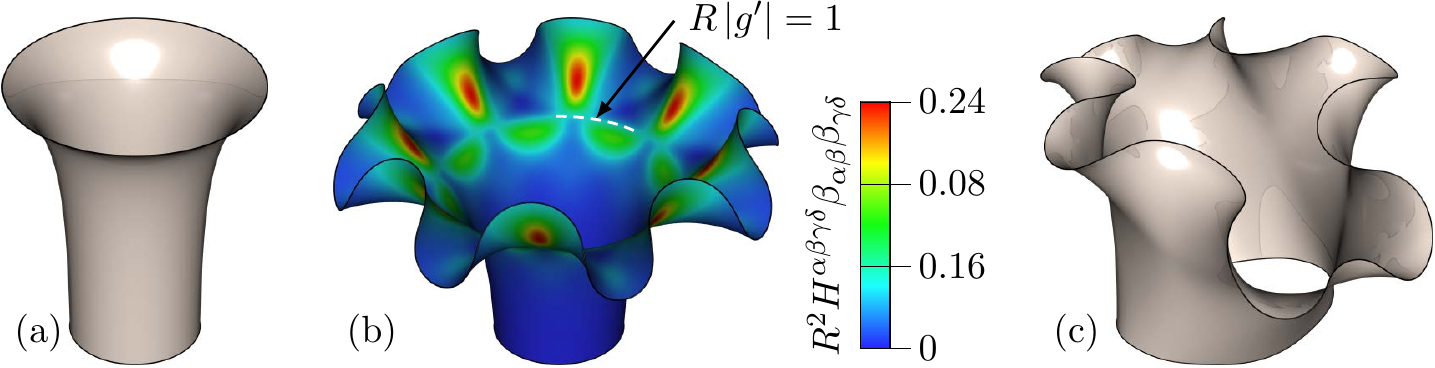}
	\caption{Boundary instability of a circumferentially growing cylinder. (a) As long as Ineq.~(\ref{eq:crit}) holds, axisymmetry is preserved. (b) Same growth field $g(z)\propto\exp(-z/l)$ as in (a), but larger prefactor. The axisymmetry is spontaneously changed to $n$-fold rotational symmetry $C_n$ ($h/R=0.02$, $l/R=1$, $n=8$). The rescaled bending energy density is shown in color, revealing the line where Ineq.~(\ref{eq:crit}) holds equally. (c) A relatively short cylinder (small $H/l$) with free boundaries also buckles away from the wrinkled edge, breaking $C_n$ symmetry further to $C_2$.}
	\label{fig:cylinder_growth}
\end{figure}

The metric profile used above is not the only one yielding wrinkled edges. When it comes to wavy flowers like certain orchids for instance, single-wavelength undulations instead of self-similar edges are not uncommon. The feature causing wrinkle cascades is the presence of a non-constant geometric length scale defined by $l_{\mathrm{geo}}(z)=-g(z)/g'(z)$ \cite{SRS07}. The following families of growth fields will thus produce similar boundary instabilities:
\begin{alignat}{3}
g_{l,p}(z) \enskip&\propto\enskip \left(1+\frac{z}{p\,l}\right)^{-p}, \qquad &(&l>0,\;p>0,\;z\geq 0),\label{eq:gbio}\\
g_{l,p}(z) \enskip&\propto\enskip \left(1-\frac{z}{p\,l}\right)^p, \qquad &(&l>0,\;p>1,\;0\leq z\leq p\,l),\label{eq:poly}
\end{alignat}
On the other hand, an exponential growth field
\begin{equation}
g_l(z) \propto\exp\left(-\frac{z}{l}\right), \qquad (l>0,\;z\geq 0)\label{eq:gexp}
\end{equation}
yields only a single wavelength \cite{MSSR03} because $l_{\mathrm{geo}}(z)\equiv l$ in this case. On some flowers, these undulations may be forced to integral wavenumbers $n$ by angular periodicity of a single petal. Let's hence significantly increase the characteristic length $l$ and thickness $h$ such that only a single wavelength $\lambda$ prevails even for growth in the form of Eqs.~(\ref{eq:gbio},\ref{eq:poly}), and let's consider a cylinder with height $H$ and radius $R$ instead of a flat plate. This change in geometry delivers dramatic consequences: A thin cylindrical sheet growing in circumferential direction according to $g(z)$, where $z$ is the cylinder axis, only breaks its axisymmetry if growth leads to a circumference that changes faster than the sheet's metric can account for \cite{M03}. (Note that the excluded linear case $p=1$ of Eq.~(\ref{eq:poly}) produces the excess cone \cite{DBA08, SWAMH10} in the limit $R\rightarrow 0$, which doesn't wrinkle at the boundary because $g_{l,1}''(z)\equiv 0$.) A direct consequence of the Gauss--Bonnet theorem is that the axisymmetry is preserved as long as 
\begin{equation}
\left|R\,\frac{\mathrm dg}{\mathrm dz}(z)\right|\leq 1,\quad 0\leq z\leq H,
\label{eq:crit}
\end{equation}
and broken otherwise. The origin for this instability is the (non)existence of \textit{embeddings} of the surface: According to Gauss's Theorema Egregium, the creation or elimination of Gaussian curvature must be accompanied by in-plane stretching, which is traded for out-of-plane buckling if the sheet is sufficiently thin (see Fig.~\ref{fig:cylinder_growth}).

\begin{figure}[!htbp]
	\centering
	\includegraphics{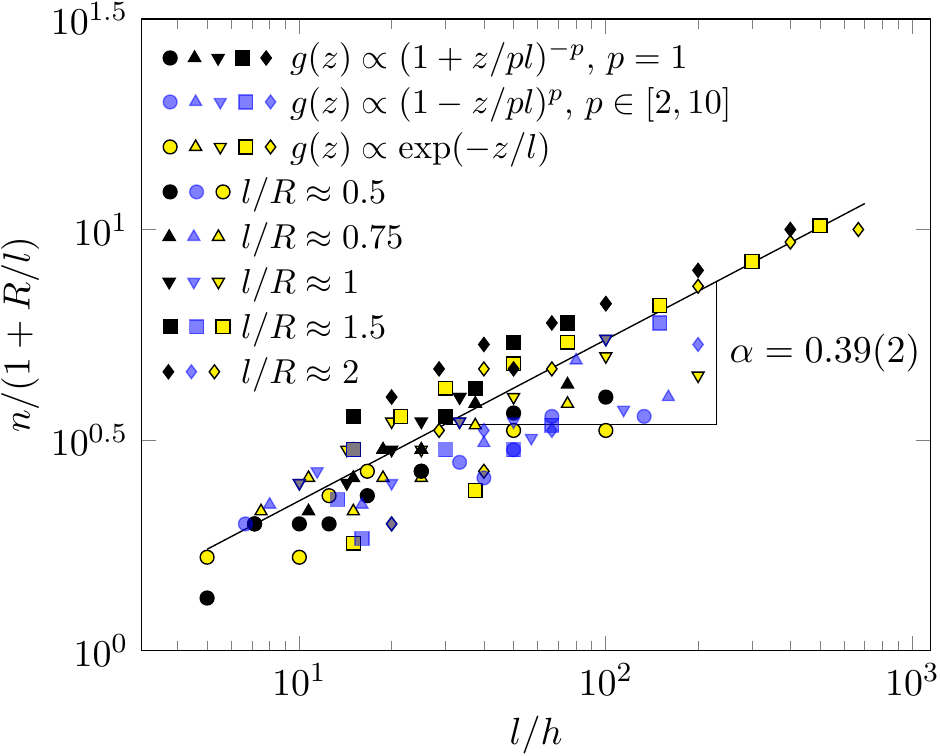}
	\caption{Scaling of the wrinkle number $n$ along the edge of a circumferentially growing cylinder. The mode with the lowest energy is the integer $n$ closest to the power law (\ref{eq:scaling_n}), but excited modes also randomly occur and are metastable (data points lying significantly above or below the straight line).}
	\label{fig:scalingplot}
\end{figure}

How does the number $n$ of boundary waves scale with $l,h,R$ when axisymmetry is broken, and is it universal for all positive, monotonically decreasing and strictly convex growth profiles satisfying
\begin{equation}
\lim_{z\to 0}l_{\mathrm{geo}}(z) = l\;?
\label{eq:char_length}
\end{equation}
Since the preferred wavelength $\lambda$ is a local feature independent of global geometry and topology (independent of $R$), we may use the ansatz \cite{SRS07}
\begin{equation}
\lambda \sim h^{\alpha}\, l_{\mathrm{geo}}^{1-\alpha},\qquad\mathrm{i.e.,}\qquad\frac{\lambda}{h} \sim \left(\frac{l_{\mathrm{geo}}}{h}\right)^{1-\alpha}.
\label{eq:lambda_ansatz}
\end{equation}
On the other hand, geometry implies that
\begin{equation}
\lambda = \frac{2\pi R}{n} \left(1+g(0)\right).
\label{eq:lambda_cyl}
\end{equation}
since $z=0$ is where Ineq.~(\ref{eq:crit}) is violated first, given that $g'<0$ and $g''>0$. After combining Eqs.~(\ref{eq:crit}\textendash\ref{eq:lambda_cyl}), one thus finds a scaling for the number of wrinkles
\begin{equation}
n \sim \left(1+\frac{R}{l}\right)\left(\frac{l}{h}\right)^{\alpha}.
\label{eq:scaling_n}
\end{equation}
Up to the first term, which accounts for the mean curvature of the cylinder, this coincides with the scaling law reported in \cite{AB03} for the wrinkling hierarchies in initially flat sheets, where $\alpha=2/5$ is found for the family $g_l(z) \propto 1/(1+z/l)$. Our numerical data, which best fits Eq.~(\ref{eq:scaling_n}) with $\alpha=0.39(2)$, shows that the wrinkles studied here fall in the same category, see Fig.~\ref{fig:scalingplot}. Moreover, the data collapse of all employed growth profiles on a single line indicates that the scaling is universal in this respect. For this numerical study, we used
\begin{equation}
{\bf F}_{\mathrm g} = \mathrm{diag}\big(1,1+g(z),1\big)
\end{equation}
in cylindrical coordinates $(r,\varphi,z)$ with various growth profiles proportional to Eqs.~(\ref{eq:gbio}-\ref{eq:gexp}), and with slowly increasing proportionality prefactors. A movie showing such spontaneous wrinkling for different $n$ is provided in the supplementary material.

\section{Confined Growth and Crumpling}

Many numerical simulations of thin sheets getting folded and crumpled inside of shrinking hollow spheres have been carried out recently \cite{KW97,VG06,TAT08,TAT09}. An important result is that the high bulk stiffness of crumpled sheets is due to a network of vertices and lines carrying large mean curvature and bending energy \cite{LGLMW95}. But what if instead of externally forced compression, the thin sheet intrinsically grows inside of a fixed spatial container? Are the processes that crumple a plant leaf or petal growing inside a bud the same as for a piece of foil crumpled by hand? Indeed they are in the elastic limit, as the following simulation demonstrates.

\begin{figure}[!htbp]
	\centering
	\includegraphics{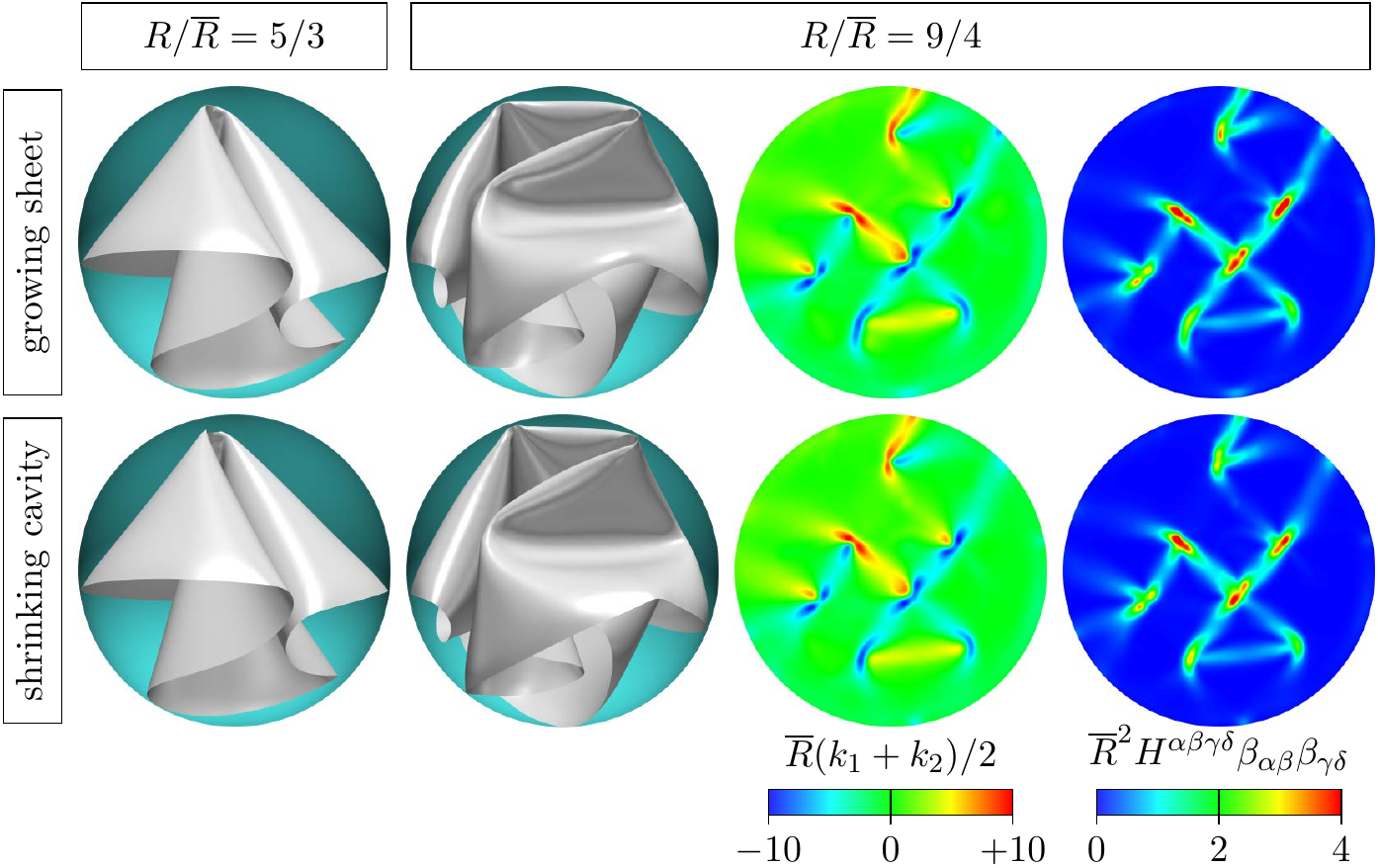}
	\caption{Comparison between shrinking confinement (bottom row) and growth in static confinement (top row), showing that the two processes are equivalent. $k_1$ and $k_2$ are the principal curvatures of the middle surface.}
	\label{fig:shrink_grow}
\end{figure}

A thin circular sheet (radius $R$) is placed inside of a spherical cavity (radius ${\overline R}$). In the first setup, the container is shrunk, folding and crumpling the sheet into a ball of the size of the container. In the second setup, the container sustains its size while the sheet undergoes uniform isotropic growth, both in plane and in thickness. The only important parameter for this problem is the F\"oppl--von K\'arm\'an number $\gamma\sim(\overline{R}/h)^2 = 10^4$. Equivalent time scales are obtained by shrinking the sphere according to $\overline{R}(t)=R/(1+g(t))$, where $g(t)=\lambda t$ is the growth factor of the growing sheet. The growth rate $\lambda$ is chosen small enough to keep inertial effects negligible. We add repulsive contact forces penalizing volumetric overlap between any two pieces of the sheet. Initially, both sheets buckle to form a developable cone with a single vertex (see Fig.~\ref{fig:shrink_grow}, first column) that starts to nucleate at ${\overline R}/R\approx 0.53$. The emerging ridge network of focused mean curvature (third column) and bending energy (fourth column) is the same in both scenarios. Compared to similar measurements \cite{TAT08}, the cross correlation $r=0.89$ of the mean curvature ridge patterns is very high. The differences are only local and of the order of the mesh resolution. A movie showing the crumpling process is contained in the supplementary material.

\section{Conclusions}

Simulating the plastic growth of thin sheets can be numerically demanding. The finite element method, being perhaps the best-suited technique for anisotropies, lacked an expedient $C^1$-continuous discretization, which is indispensable from a theoretical viewpoint, until subdivision surfaces were ported to it. We have extended the Kirchhoff--Love theory by arbitrary volumetric in-plane growth \cite{VSJWH13} and used the Loop subdivision surface paradigm to build a highly flexible and efficient numerical tool for the simulation of nonlinear thin sheet mechanics. Requiring no rotational variables, a thin sheet representation of this kind is remarkably simple to implement and superior to traditional approaches in terms of computational costs. A series of example simulations were carried out to demonstrate these strengths. In particular, we have quantified the self-similarity and scaling of wrinkles along plastically stretched free edges of thin sheets, and found that the problem of a growing confined sheet is equivalent to the narrowing confinement of a static sheet.

\section*{Acknowledgments}

Financial support support from the ETH Research (ETHIIRA) Grant No.~ETH-03~10-3 as well as from the European Research Council (ERC) Advanced Grant No.~319968-FlowCCS is gratefully acknowledged.

\end{document}